\documentclass[letterpaper,english,floatfix,pre,twocolumn]{revtex4}
\usepackage[]{fontenc}
\usepackage[latin9]{inputenc}
\usepackage{babel}

\usepackage{amsmath}
\usepackage{graphicx}
\usepackage{esint}
\usepackage[unicode=true, 
 bookmarks=true,bookmarksnumbered=false,bookmarksopen=false,
 breaklinks=true,pdfborder={0 0 0},backref=false,colorlinks=false]
 {hyperref}
\hypersetup{pdftitle={MF theory of a plastic network of I&F neurons},
 pdfauthor={Chun-Chung Chen}}
 
\makeatletter

\providecommand{\tabularnewline}{\\}

\@ifundefined{textcolor}{}
{%
 \definecolor{BLACK}{gray}{0}
 \definecolor{WHITE}{gray}{1}
 \definecolor{RED}{rgb}{1,0,0}
 \definecolor{GREEN}{rgb}{0,1,0}
 \definecolor{BLUE}{rgb}{0,0,1}
 \definecolor{CYAN}{cmyk}{1,0,0,0}
 \definecolor{MAGENTA}{cmyk}{0,1,0,0}
 \definecolor{YELLOW}{cmyk}{0,0,1,0}
 }

\makeatother

\begin{document}

\title{Mean-field theory of a plastic network of integrate-and-fire neurons}

\author{Chun-Chung Chen}

\author{David Jasnow}

\affiliation{Department of Physics and Astronomy, University of Pittsburgh, Pittsburgh,
Pennsylvania 15260, USA}

\date{November 13, 2009}
\begin{abstract}
We consider a noise driven network of integrate-and-fire neurons.
The network evolves as result of the activities of the neurons following
spike-timing-dependent plasticity rules. We apply a self-consistent
mean-field theory to the system to obtain the mean activity level
for the system as a function of the mean synaptic weight, which predicts
a first-order transition and hysteresis between a noise-dominated
regime and a regime of persistent neural activity. Assuming Poisson
firing statistics for the neurons, the plasticity dynamics of a synapse
under the influence of the mean-field environment can be mapped to
the dynamics of an asymmetric random walk in synaptic-weight space.
Using a master-equation for small steps, we predict a narrow distribution
of synaptic weights that scales with the square root of the plasticity
rate for the stationary state of the system given plausible physiological
parameter values describing neural transmission and plasticity. The
dependence of the distribution on the synaptic weight of the mean-field
environment allows us to determine the mean synaptic weight self-consistently.
The effect of fluctuations in the total synaptic conductance and plasticity
step sizes are also considered. Such fluctuations result in a smoothing
of the first-order transition for low number of afferent synapses
per neuron and a broadening of the synaptic weight distribution, respectively.
\end{abstract}
\maketitle

\section{Introduction}

The brains of living animals are perhaps the most complex organs one
can find. These are networks of neurons formed through the connecting
synapses, and their proper functioning is crucial to survival. Theoretical
studies of the dynamics of neural networks have contributed to our
understanding of how these networks might function \cite{vogels_neural_2005,rabinovich_dynamical_2006}.
Recent progress in the study of synaptic plasticity is opening up
new opportunities for understanding how these networks can form \cite{abbott_synaptic_2000,roberts_spike_2002,dan_spike_2004}.
One of the most important aspects of a functional neural network is
that the strength of its connections can change in response to the
history of its activities, that is, it can learn from experience \cite{amit_dynamics_1997,malenka_long-term_1999,lynch_long-term_2004,collingridge_receptor_2004}.
Among the theories of neural learning, the best known, Hebbian theory,
states that when a neuron partakes in the firing of another neuron,
its ability of doing so increases \cite{hebb_organization_1949}.
An oversimplified version of the theory states that {}``cells that
fire together, wire together'' which is generally adequate for rate-based
views of neurons \cite{gerstner_spiking_2002}. However, the actual
activities of neurons are described by {}``spikes'', that is, action
potentials that are produced when the state of a neuron meets certain
criteria, for example, when its membrane potential reaches a threshold
value \cite{izhikevich_dynamical_2006}. It was discovered that the
exact timing of these spikes plays a great role in determining the
synaptic plasticity, adding an element of causation to the correlation
requirement of the learning process. That is, the presynaptic neuron
must fire before the postsynaptic neuron for the former to actually
take part in firing the later. In fact, it was discovered that a sharp
change from potentiation to depression of the synaptic efficacy can
occur as the firing time of the postsynaptic neuron precedes that
of the presynaptic one \cite{bi_synaptic_2001,roberts_spike_2002,dan_spike_2004,caporale_spike_2008}.

An important feature of Hebbian learning is the presence of positive
feedback. That is, the stronger a synapse is, the more likely that
it will be potentiated. Often in simulated plastic networks, this
presents a possibility of runaway synaptic weights, which are often
curbed through the introduction of cutoffs or other devices. In some
schemes, the potentiation of a synapse is suspended when its weight
exceeds a preset cutoff value \cite{amarasingham_predictingdistribution_1998,song_competitive_2000,rubin_equilibrium_2001,toyoizumi_optimality_2007}.
While such cutoffs might be justified by physiological limitations
of the cells, in simulation studies they often result in a pile-up
of synaptic weights distributed near the cutoffs that is not observed
in real biological neural networks \cite{blum_model_1996,gerstner_neuronal_1996,amarasingham_predictingdistribution_1998,kempter_hebbian_1999,song_competitive_2000,van_rossum_stable_2000}.
Runaways or pile-ups can often be mitigated by softening the cutoff
\cite{kistler_modeling_2000,gerstner_spiking_2002,toyoizumi_optimality_2007}.
Alternatively, it has been proposed from experimental observations
that the potentiation and depression processes could be \emph{asymmetric},
such that potentiation is an additive process to the current synaptic
weight while depression is a multiplicative process. Such a mechanism
was shown to produce a uni-modal distribution of synaptic weights
free of pile-ups among the many afferent synapses of a \emph{single
neuron} when the inputs are driven by Poisson spiking sources \cite{van_rossum_stable_2000}.

Besides viewing brains \emph{microscopically} as networks of neurons
as discussed above, efforts aimed at understanding the structure and
function of a living brain also include studies at the \emph{macroscopic}
scale, for example of the interactions between different anatomical
regions or even of the role played by the entirety of the brain as
a vital organ \cite{angevine_principles_1981,hebb_organization_1949}.
The contrast of the two scales is very analogous to the study of condensed
matter systems, which has faced similar challenges of bridging our
understanding of the microscopic with our observation of the macroscopic.
Among various tools employed by theorists studying condensed matter
systems, mean-field theory has been proven valuable in that it often
allows one to obtain a quick grasp of what macroscopic states of the
systems can be expected from the set of microscopic mechanisms they
follow\cite{huang_statistical_1988}. While it is well understood
that mean-field theories could fail to predict correct scaling behavior
of systems in critical states, where fluctuations and long-range correlations
are important, they generally are adequate (and often the first step
in a systematic procedure) in qualitatively describing the system
in stable phases, where fluctuations are of limited range, and useful
in revealing the structure of the phase spaces of the systems. The
later is especially desirable for biological systems where large numbers
of empirical parameters and, consequently, vast phase spaces are often
involved in microscopic models of the systems.

In the current study, we consider a network of integrate-and-fire
neurons\cite{burkitt_review_2006} driven by Poisson noise of fixed
frequency for all neurons. The interaction between the neurons is
taken to follow the neural transmission model proposed by Tsodyks,
Uziel, and Markram \cite{tsodyks_synchrony_2000} (the TUM model),
which can account for the saturation effect of neural transmitter
and the short-term depression of synaptic conductance. While our approach
can handle alternative choices, the synaptic weights between the neurons
are allowed to evolve following the spike-timing-dependent plasticity
rules proposed by Bi and coworkers \cite{bi_synaptic_1998,van_rossum_stable_2000}.
A mean-field theory is used to determine the self-consistent average
firing rate of the neurons. The noise driven firings dominate the
small synaptic weight regime, while self-sustaining firing activity
is triggered in the large synaptic weight regime. As the mean synaptic
weight of the network is varied, the mean-field theory predicts a
hysteresis for the transition between regimes of noise dominance and
self-sustaining activity. Within the mean-field environment, assuming
the neurons are firing with Poisson statistics, the dynamics of a
single synapse can be viewed as a random walk process in synaptic-weight
space. Under the small-jump assumption, we use the master equation
to calculate the drift and diffusion coefficient for the random walk.
The resulting Fokker--Planck equation allows us to predict the stationary
synaptic-weight distribution of the process and close the self-consistency
with the requirement that the stationary distribution reproduces the
mean synaptic weight characterizing the mean-field environment. For
the plasticity rules and ranges of the parameters we considered, the
synaptic weights form a narrow distribution, having a width proportional
to square root of the plasticity rate. Finally, we extend the mean-field
approximation to include the effect of fluctuations in the synaptic
conductance as well as the variations of jump sizes in the random
walk of synaptic weights.

Of course, despite our mention of brains of living creatures in our
introduction, the current analysis cannot hope to address issues of
dynamics involved in such a venue for myriad reasons. Among others,
here we imagine homogeneous, stationary networks, which is surely
not the case in brain. However, there are continuing, revealing experiments
on \emph{cultured }neural\emph{ }networks with perhaps several hundred
individual neurons in which all-to-all coupling is not an unreasonable
approximation or starting point \cite{bi_synaptic_1998,segev_observations_2001,shefi_morphological_2002,beggs_neuronal_2003,lai_growth_2006}.
This class of experiments represents an important step and will continue
to provide important insights into the behavior of more complicated
networks. Our aims are to improve the understanding of the stationary,
statistical properties of such plastic network and ultimately address
dynamical behavior during formation. Furthermore, cultured networks
are on a scale approachable by the modeling, numerical and analytic
work such as presented here. By analogy with a variety of familiar
statistical mechanical models, sufficiently large homogeneous networks
in which the number of {}``neighbors'' of a particular element grows
proportionally to the number of elements in the thermodynamic limit,
are expected to be well-described by the type of mean-field analysis
employed in this study (see, \emph{e.g.}, chapter 3 of \cite{baker_quantitative_1990}).
In a separate publication we will present results of extensive numerical
simulations on integrate-and-fire (and other representative neuron
model) networks, the results of which can be put in the proper perspective
via comparison with the calculations presented here and with results
of\emph{ in vitro }experiments. 

The remainder of this paper is laid out as follows. In Section~\ref{sec:Model},
we describe the model of the plastic network, which consists of (i)
the dynamics of the membrane potential, (ii) the model of synaptic
transmission, and (iii) the plasticity rules. In Section~\ref{sec:Mean-field-approximation},
we apply the mean-field method to a network with fixed synaptic weights
and determine the state of the system through response functions deduced
from the dynamics of neuron and synaptic transmission under a given
mean synaptic weight. In Section~\ref{sec:Random-walk-in-w-space},
we map synaptic plasticity to a random walk process in synaptic-weight
space, determine the stationary synaptic-weight distribution, and
close the self-consistency in the mean synaptic weight of the network.
In Section~\ref{sec:Correction-to-MFT}, we expand the mean-field
theory to include the consideration of fluctuations in total synaptic
current of a neuron and fluctuations in step sizes of synaptic weight
changes. Finally, we summarize and conclude in Section~\ref{sec:Summary-and-conclusion}.

\section{Model\label{sec:Model}}

We consider a noise-driven plastic network of integrate-and-fire neurons.
The neurons are coupled using the TUM model of neural transmission
\cite{tsodyks_synchrony_2000,volman_calcium_2007}, described below.
The noise is modeled by randomly forced firing of the neurons following
Poisson statistics at a fixed frequency.

\subsection{Integrate-and-fire neuron}

The integrate-and-fire model is a single-compartment neuron model
where the state of a neuron $i$ is described by a membrane potential
$V_{i}$. The dynamics of the membrane potential follows the differential
equation of a leaky integrator \cite{dayan_theoretical_2001}

\begin{equation}
\tau_{m}\frac{dV_{i}}{dt}=V_{0}-V_{i}+R_{m}I_{\mathrm{syn}},\label{eq:membrane-potential}\end{equation}
where $\tau_{m}$ is the leak time for the membrane charge, which
is given by the product of the total membrane capacitance $C_{m}$
and resistance $R_{m}$, while $V_{0}$ is the resting potential when
the neuron is in the quiescent state. The total synaptic current $I_{\mathrm{syn}}$
is a sum over each \emph{afferent} synapse $j$ for the neuron $i$
\begin{equation}
I_{\mathrm{syn}}=\frac{1}{R_{m}}\sum_{j}w_{j,i}Y_{j,i}\left(\mathcal{R}_{j,i}-V_{i}\right),\end{equation}
 where, for the synapse connecting neuron $j$ to $i$, $w_{j,i}$
is the synaptic weight, $\mathcal{R}_{j,i}$ is the reversal potential
for the ion channels, and $Y_{j,i}$ is the fraction of active transmitters.
The dimensionless synaptic weight $w_{j,i}$ can be interpreted as
the maximum synaptic conductance, achieved when $Y_{j,i}=1$, measured
in units of the membrane conductance $R_{m}^{-1}$ of the neuron.
In the current study, we consider cases in which there is only one
type of ion channel at all synapses so that they share the same reversal
potential $\mathcal{R}_{j,i}=\mathcal{R}$. In these cases, the difference
of membrane potential from the reversal potential can be factored
out and Eq.~\eqref{eq:membrane-potential} becomes \begin{equation}
\tau_{m}\frac{dV_{i}}{dt}=V_{0}-V_{i}+G_{i}\left(\mathcal{R}-V_{i}\right),\label{eq:membrane-with-conductance}\end{equation}
where \begin{equation}
G_{i}\equiv\sum_{j}w_{j,i}Y_{j,i}\label{eq:total-conductance}\end{equation}
is the total synaptic conductance (in units of $R_{m}^{-1}$).

In this model, a neuron fires when its membrane potential reaches
a threshold value, $V_{\mathrm{th}}$. Then, its membrane potential
drops immediately to a reset value, $V_{\mathrm{r}}$. The action
potential of the integrate-and-fire model is assumed to be instantaneous
and is not modeled explicitly. The spike train produced by the neuron
$i$ is defined as the function \begin{equation}
S_{i}\equiv\sum_{n}\delta\left(t-t_{i,n}\right)\label{eq:spike-train}\end{equation}
 where $t_{i,n}$ is the time when the neuron $i$ fires for the $n$-th
time.

\subsection{Tsodyks--Uziel--Markram model of neural transmission}

The fractions $Y_{j,i}$ of the active transmitters are described
by the TUM model \cite{tsodyks_synchrony_2000} of neural transmission,
where the transmitters are distributed in three states: {}``active'',
with the fraction $Y$; {}``inactive'', with the fraction $Z$;
and {}``ready-to-release'', with the fraction $X$. For a synapse
with presynaptic neuron $j$ and postsynaptic neuron $i$, these fractions
follow the dynamics \cite{tsodyks_synchrony_2000} \begin{eqnarray}
\frac{dX_{j,i}}{dt} & = & \frac{Z_{j,i}}{\tau_{R}}-uS_{j}X_{j,i}\nonumber \\
\frac{dY_{j,i}}{dt} & = & -\frac{Y_{j,i}}{\tau_{D}}+uS_{j}X_{j,i}\label{eq:TUM-model}\\
\frac{dZ_{j,i}}{dt} & = & \frac{Y_{j,i}}{\tau_{D}}-\frac{Z_{j,i}}{\tau_{R}},\nonumber \end{eqnarray}
where $\tau_{D}$ is the decay time of active transmitters to the
inactive state, $\tau_{R}$ is the recovery time for the inactive
transmitters to the ready-to-release state, and $u$ is the fraction
of ready-to-release transmitters that is released to the active state
by each presynaptic spike. With the conservation rule \begin{equation}
X_{j,i}+Y_{j,i}+Z_{j,i}=1,\label{eq:transmitter-conservation}\end{equation}
there are two independent variables per synapse. Since the multiplying
factor of the spike train $S_{j}$ in the dynamics of $X_{j,i}$ depends
on $X_{j,i}$ itself, which is discontinuous when there is a $\delta$-function
in $S_{j}$ due to a spike at given time $t$, we must specify how
the value of $X_{j,i}$ should be evaluated at the time of the spike.
Consistent with the TUM dynamics, the values of the factors multiplying
$S_{j}$ at the discontinuities are to be evaluated immediately before
the discontinuities.

The TUM model is very flexible. Since the variable $Y$ represents
a fraction in the TUM model, its value can saturate. As a result,
an increase in the firing of the presynaptic neuron can only produce
a less-than-linear increase of the active transmitters. However, a
linearity of active transmitters on the firing rate can be recovered
in the $u\rightarrow0$ limit, maintaining the average level of the
synaptic conductance by keeping the product $w_{j,i}u$ constant.
Additionally, the presence of the inactive state, $Z$, mimics the
short-term depression of neural transmission, where repeated firings
of the presynaptic neuron over a short period of time will reduce
the maximum value attainable by $Y$ temporarily through depositing
the transmitters into the inactive state before their recovery. Finally,
manipulating $\tau_{R}/\tau_{D}$ can reproduce a model with a single,
saturating {}``species'' Y (see below).

Since the dynamics \eqref{eq:TUM-model} depend only on the spike
train of the presynaptic neuron $j$, without external disturbance
all the \emph{efferent} synapses of a neuron should have the same
values of transmitter fractions. Thus, we can drop the subscript $i$
from dynamics \eqref{eq:TUM-model} and regard these fractions as
the properties of the presynaptic neuron $j$. Such simplification
is not applicable for transmitter dynamics that depend, for example,
on the postsynaptic neuron, that have synaptic-weight-dependent parameters,
or synapse-dependent noise. These complications are not considered
here.

\subsection{Noise}

While the integrate-and-fire and TUM dynamics are both deterministic,
we model the stochasticity of the network with additional noise driven
firing events following Poisson statistics with the frequency $\lambda_{N}$
for each neuron. The noise driven firings are treated the same way
as threshold firings, that is, the membrane potentials are brought
instantaneously to the reset value $V_{\mathrm{r}}$ and the firing
times are included in the spike trains \eqref{eq:spike-train} of
the transmitter dynamics \eqref{eq:TUM-model}.

\subsection{Spike-timing-dependent plasticity}

It has been observed experimentally that the change of synaptic efficacy
depends on the precise timing between the presynaptic and postsynaptic
spikes \cite{bi_synaptic_2001}. When a presynaptic spike precedes
a post-synaptic spike, following van Rossum, Bi, and Turrigiano \cite{van_rossum_stable_2000},
we take the synapse to be potentiated by an amount proportional to
$e^{-\Delta t/\tau_{A}}$ where $\Delta t$ is the timing difference
between the two spikes and $\tau_{A}$ is the size of the potentiation
time window. Similarly when a postsynaptic spike precedes a presynaptic
spike, the synapse is taken to be depressed by an amount proportional
to $e^{-\Delta t/\tau_{B}}$ where $\tau_{B}$ is the size of the
depression time window \cite{bi_synaptic_1998,van_rossum_stable_2000}.
One reasonable mechanism for the cell to determine the interval between
spikes is to imagine the cell evaluates the concentration of a decaying
chemical species released at the first spike \cite{pfister_triplets_2006}.
Following such a mechanism, the synaptic changes proposed in \cite{van_rossum_stable_2000}
can be produced for the synaptic weight $w_{j,i}$ by the dynamical
equation \begin{equation}
\frac{dw_{j,i}}{dt}=\Delta A_{j}S_{i}-rw_{j,i}B_{i}S_{j},\label{eq:weight-dynamics}\end{equation}
where $\Delta$ is the potentiation constant for causal spike pairs,
$r$ is the depression factor for anti-causal spike pairs, while $A$s
and $B$s are the amplitudes of such (assumed) chemical species controlling
the magnitude of potentiation and depression. These amplitudes are
assumed to follow the dynamics \cite{pfister_triplets_2006} \begin{equation}
\frac{d\sigma_{j}}{dt}=-\frac{\sigma_{j}}{\tau_{\sigma}}+f_{\sigma}S_{j},\;\sigma=A,B\label{eq:chemical-decay}\end{equation}
where different choices of the spike increments $f_{\sigma}$ lead
to different schemes of spike pairing. For example, $f_{\sigma}=1$
leads to {}``all-to-all'' pairing of presynaptic and postsynaptic
spikes while $f_{\sigma}=1-\sigma$ leads to {}``nearest-neighbor''
pairing of spikes \cite{pfister_triplets_2006}. We note the stochastic
differential equations in Eqs.~\eqref{eq:weight-dynamics} and \eqref{eq:chemical-decay}
above should follow Itô's interpretation \cite{van_kampen_stochastic_2007}.
That is, when the value of $\sigma$ jumps because of a spike in $S$,
the value of $f_{\sigma}$, which determines the size of the jump,
is evaluated immediately before the jump. The effects of the two choices
of $f_{\sigma}$ are identical when spike pairs are far and apart
but differ when possible spike pairs are close compared with the decay
time $\tau_{\sigma}$ of the respective timing chemicals. With the
choice $f_{\sigma}=1$, the values of $A$ or $B$ can increase linearly
with the frequency of spikes without limit. On the other hand, for
the choice $f_{\sigma}=1-\sigma$, the value of $\sigma$ will never
exceed 1, its value immediately after a single spike. It is reasonable
to expect, in a realistic setting, that the values of $A$ or $B$
will increase with increasing spike rates, but they should saturate
and be limited to some physiologically determined maximum values with
further increase in spike rates. Assuming spike increments $f_{\sigma}=u_{\sigma}\left(1-\sigma\right)$
with $u_{\sigma}<1$ can produce this behavior. The parameter $u_{\sigma}$
can be interpreted as the mean fraction of chemicals released at each
presynaptic firing relative to the amount that can possibly be produced
without $\sigma$ exceeding its maximum value of $1$.

Additionally, certain effects of fatigue can also be expected in synaptic
plasticity. \cite{froemke_contribution_2006} That is, while $\sigma$
($=$$A$ or $B$) can saturate when the spike rate increases, the
saturation value itself can be expected to decrease when a high spiking
rate is sustained for an extended period of time \cite{tsodyks_synchrony_2000,volman_calcium_2007}.
Such effect can be modeled by assuming an additional inactive state
$I_{\sigma}$, which takes up a fraction of $\sigma$ that can be
produced for each spike, so that the increment becomes $f_{\sigma}=u_{\sigma}\left(1-\sigma-I_{\sigma}\right)$.
One can assume simple dynamics for $I_{\sigma}$, given by \begin{equation}
\frac{dI_{\sigma}}{dt}=\frac{\sigma}{\tau_{I\sigma}}-\frac{I_{\sigma}}{\tau_{R\sigma}}\label{eq:inactive_dynamics}\end{equation}
so that $I_{\sigma}$ is fed by the presence of $\sigma$ with the
rate $\tau_{I\sigma}^{-1}$ and decays with a recovery time constant
$\tau_{R\sigma}$. One expects that the rate of $\sigma$'s feeding
into the inactive state {[}first term on the right of Eq.~\eqref{eq:inactive_dynamics}{]}
should be less than the rate of its own decay {[}first term on the
right of Eq.~\eqref{eq:chemical-decay}{]}, so we require the condition
$\tau_{I\sigma}\geq\tau_{\sigma}$.

The considerations outlined above conspire to produce a variant of
the TUM model similar to what we have used for the neural transmitters
apart from possibly different values of the constants involved. In
the current simulation study of plastic integrate-and-fire networks,
we retain the qualitative structure of the system but simplify by
assuming $\tau_{\sigma}=\tau_{I\sigma}=\tau_{D}$, $\tau_{R\sigma}=\tau_{R}$,
and $u_{\sigma}=u$ so that the same values of $Y_{i}$ calculated
for the neural transmitter in the TUM model can be used as surrogates
for $A_{i}$ and $B_{i}$ in the synaptic plasticity yielding \begin{equation}
\frac{dw_{j,i}}{dt}=\Delta Y_{j}S_{i}-rw_{j,i}Y_{i}S_{j}.\label{eq:simplified stdp}\end{equation}
These simplifications reduce the amount of computation required without
sacrificing qualitative and semi-quantitative analysis.

To summarize, our system of equations reduces to Eqs.~\eqref{eq:membrane-with-conductance}--\eqref{eq:transmitter-conservation}
and \eqref{eq:simplified stdp}. The {}``control parameter'' in
subsequent analysis will turn out to be \begin{equation}
w^{\star}\equiv\frac{\Delta}{r},\label{eq:wstar-definition}\end{equation}
 which, when fixed, leaves the depression factor $r$ as an overall
control of rate of plasticity.

\section{Mean-field approximation\label{sec:Mean-field-approximation}}

\subsection{Single neuron response function}

To formulate a mean-field approximation, we assume that the firing
rates of the neurons are given by the same mean-field value $\bar{\lambda}$
and that all the synapses have the same synaptic weight $\bar{w}$.
The total synaptic conductance $G$ of a singled-out postsynaptic
neuron is a function of time and jumps whenever there is a firing
of a presynaptic neuron and subsequently decays with the time constant
$\tau_{m}$ as in Eq.~\eqref{eq:membrane-with-conductance}. Consider
the limit of large number $K$ of afferent synapses per neuron while
keeping the product $\bar{w}K$ constant. Then, the total frequency
of all presynaptic firings is $K\bar{\lambda}\rightarrow\infty$.
In this limit, the jump sizes of $G$, being proportional to $\bar{w}$,
approach $0$ while the rate of making the jumps diverges. For fixed
$\bar{\lambda}$ and $\bar{w}K$, we may thus ignore fluctuations
and consider $G$ as a constant over time. 

\begin{figure}
\begin{centering}
\includegraphics[scale=1.3]{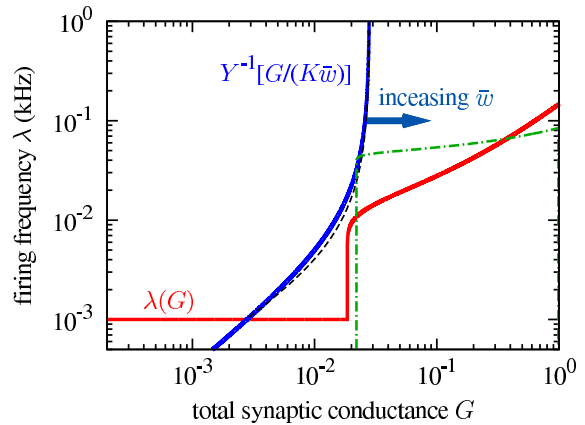}
\par\end{centering}

\caption{Firing frequency $\lambda\left(G\right)$ of an integrate-and-fire
neuron under constant total synaptic conductance $G$ given by Eq.~\eqref{eq:lambda-function}.
Also shown are mean-field active synaptic transmitter fractions $Y\left(\lambda\right)$
(for $K=31$, $\bar{w}=0.01$) driven by a Poisson spike train of
frequency $\lambda$ given by \eqref{eq:synapse-response-function}.
The dashed line represents a numerically calculated correction to
$Y\left(\lambda\right)$ due to the periodic firing of the neuron
when driven by constant total synaptic conductance $G$. The parameters
are listed in Table~\ref{tab:Values-of-parameters}. The arrow indicates
that the curve shifts to the right for increasing $\bar{w}$. The
dot-dashed line is the neuron response function numerically calculated
for the Morris--Lecar model used in \cite{volman_calcium_2007} with
a slightly higher background current. \label{fig:Firing-frequency-on-G} }

\end{figure}

For a constant total synaptic conductance $G$, the firing frequency
of an integrate-and-fire neuron can be solved exactly by setting the
initial membrane potential to $V\left(0\right)=V_{\mathrm{r}}$, the
reset value, and finding the time $\tau$ it takes to reach the firing
threshold $V\left(\tau\right)=V_{\mathrm{th}}$. The solution of Eq.~\eqref{eq:membrane-with-conductance}
is given by \begin{equation}
\tau=-\frac{\tau_{m}}{G+1}\ln\left(\frac{\tilde{V}\left(G\right)-V_{\mathrm{th}}}{\tilde{V}\left(G\right)-V_{\mathrm{r}}}\right)\label{eq:IF-period}\end{equation}
when $\tilde{V}\left(G\right)\geq V_{\mathrm{th}}$, where \begin{equation}
\tilde{V}\left(G\right)\equiv\frac{V_{0}+G\mathcal{R}}{G+1}\end{equation}
 is the resting value for the membrane potential under the constant
conductance $G$. For $\tilde{V}<V_{\mathrm{th}}$, we have $\tau=\infty$
since the neuron never fires. In addition to firings due to crossing
the threshold, the period $\tau$ can be cut short by Poisson noise
of frequency $\lambda_{N}$ as the membrane potential sweeps from
$V_{\mathrm{r}}$ to $V_{\mathrm{th}}$. Given Poisson statistics,
the average interval $\bar{\tau}$ between all firings (either threshold-crossing
or noise-driven) is given by \begin{eqnarray}
\bar{\tau} & = & \int_{0}^{\tau}dt\lambda_{N}te^{-\lambda_{N}t}+\tau\int_{\tau}^{\infty}dt\lambda_{N}e^{-\lambda_{N}t}\nonumber \\
 & = & \lambda_{N}^{-1}\left(1-e^{-\lambda_{N}\tau}\right),\end{eqnarray}
and the firing frequency $\lambda$ of the neurons is given by \begin{equation}
\lambda\left(G\right)=\lambda_{N}\left[1-\left(\frac{\tilde{V}\left(G\right)-V_{\mathrm{th}}}{\tilde{V}\left(G\right)-V_{\mathrm{r}}}\right)^{\frac{\lambda_{N}\tau_{m}}{G+1}}\right]^{-1}\label{eq:lambda-function}\end{equation}
for $\tilde{V}\geq V_{\mathrm{th}}$, and by $\lambda=\lambda_{N}$
otherwise. (See Fig.~\ref{fig:Firing-frequency-on-G}.) %
\begin{table}
\caption{Values of parameter\label{tab:Values-of-parameters}s used in calculations}

\centering{}\begin{tabular}{rlccrl}
\multicolumn{2}{c}{integrate \& fire} &  &  & \multicolumn{2}{c}{TUM model}\tabularnewline
\cline{1-2} \cline{5-6} 
resting potential $V_{0}$: & -55 mV &  &  & decay time $\tau_{D}$: & 20 ms\tabularnewline
leak time $\tau_{m}$: & 20 ms &  &  & recovery time $\tau_{R}$: & 200 ms\tabularnewline
firing threshold $V_{\mathrm{th}}$: & -54 mV &  &  & release fraction $u$: & 0.5\tabularnewline
\cline{5-6} 
reset potential $V_{\mathrm{r}}$: & -80 mV &  &  & noise frequency $\lambda_{N}$: & 1 Hz\tabularnewline
reversal potential $\mathcal{R}$: & 0 mV &  &  & plasticity rate $r$: & 0.01\tabularnewline
\end{tabular}
\end{table}
 The function $\lambda\left(G\right)$ obtained in \eqref{eq:lambda-function}
characterizes the mean-field response of the integrate-and-fire network
model in the current study %
\footnote{In this article, the symbols $\lambda$ and $Y$, when used as a function,
\emph{e.g.}, $\lambda\left(G\right)$ or $Y\left(\bar{\lambda}\right)$,
always represent the response functions of the neuron and synapse
respectively. Otherwise, they are only variables standing for a firing
frequency or an active transmitter fraction.%
}, but similar functions can be obtained analytically or numerically
for different neuron models, such as, the Hodgkin--Huxley, Morris--Lecar,
or FitzHugh--Nagumo models (see, \emph{e.g.}, \cite{gerstner_spiking_2002,izhikevich_which_2004}).
(An example of a numerically calculated response function for a Morris--Lecar
neuron is included in Fig.~\ref{fig:Firing-frequency-on-G}.)

\subsection{Synapse response function}

To obtain the mean-field active transmitter fraction $\bar{Y}$ of
a neuron, we further approximate the firing of the neurons as having
Poisson statistics described by the mean-field firing rate $\bar{\lambda}$
\cite{rubin_equilibrium_2001}. Keeping $Y$ and $Z$ as the independent
variables for the TUM model, the stochastic dynamics \eqref{eq:TUM-model}
can be averaged over an ensemble of Poisson spike trains with $\left\langle S\right\rangle =\lambda$
for the presynaptic neuron to yield the functions \begin{equation}
Y\left(\lambda\right)=\frac{u\tau_{D}\lambda}{1+u\left(\tau_{D}+\tau_{R}\right)\lambda}\label{eq:synapse-response-function}\end{equation}
and $Z\left(\lambda\right)=\frac{\tau_{R}}{\tau_{D}}Y\left(\lambda\right)$
for the average active ($Y$) and inactive ($Z$) transmitter fractions.
The function $Y\left(\lambda\right)$ characterizes the mean-field
response of a synapse \footnotemark[\value{footnote}]. The value
of the active transmitter fraction for a mean-field network can be
obtained through the relation \begin{equation}
\bar{Y}=Y\left(\bar{\lambda}\right).\end{equation}

\subsection{Self-consistency condition}

The total synaptic conductance of a neuron in a mean-field network
with $K$ afferent synapses per neuron is given by \begin{equation}
\bar{G}=K\bar{w}\bar{Y}=K\bar{w}Y\left(\bar{\lambda}\right),\label{eq:transmitter-to-conductance}\end{equation}
which can be substituted into Eq.~\eqref{eq:lambda-function} to
complete the self-consistency condition closing the set of Eqs.~\eqref{eq:lambda-function},
\eqref{eq:synapse-response-function}, and \eqref{eq:transmitter-to-conductance}.
The curve for Eq.~\eqref{eq:transmitter-to-conductance} is also
shown in Fig.~\ref{fig:Firing-frequency-on-G}, and the intersections
with the neuron response function $\lambda\left(G\right)$ represent
fixed points of the network activities, $\bar{\lambda}$, given the
mean-field synaptic weight $\bar{w}$. %
\begin{figure}
\begin{centering}
\includegraphics[scale=1.3]{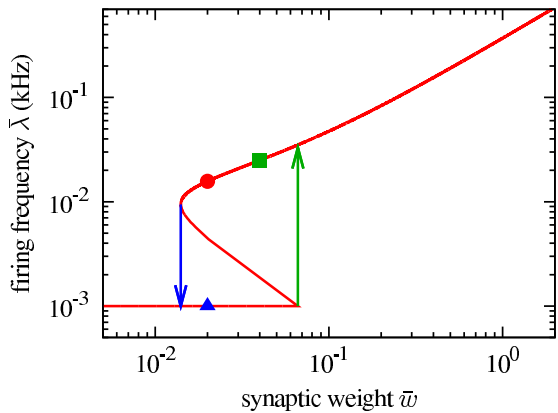}
\par\end{centering}

\caption{Mean-field firing frequency $\bar{\lambda}$ as a function of mean-field
synaptic weight $\bar{w}$ determined by the intersections of neuron
response function and synapse response function as plotted in Fig.~\ref{fig:Firing-frequency-on-G}
for $K=31$. The two arrows show the region of hysteresis where two
stable fixed points, the noise-dominating lower fixed point and the
persistently active upper fixed point, coexist. The symbols will be
used in connection with Figs.~\ref{fig:Mean-synaptic-weight} and
\ref{fig:limited-system-results}. \label{fig:MF-frequency-on-weight}}

\end{figure}
 The function $\bar{\lambda}\left(\bar{w}\right)$ is shown in Fig.~\ref{fig:MF-frequency-on-weight}.
The mean-field theory predicts a first-order phase transition and
hysteresis as the mean-field synaptic weight $\bar{w}$ is varied.
At low synaptic weight, the activities of the network are dominated
by the external noise. As the synaptic weight is increased, a pair
of new fixed points emerge at higher firing frequencies. Among the
three fixed points, only the upper and the (noise-dominated) lower
fixed points are stable. As the synaptic weight is increased further,
the lower fixed point eventually is annihilated by the unstable (middle)
fixed point leaving only the upper fixed point representing higher
firing activities.

\section{Random walk in synaptic weight space\label{sec:Random-walk-in-w-space}}

\begin{figure}
\begin{centering}
\includegraphics[scale=0.4]{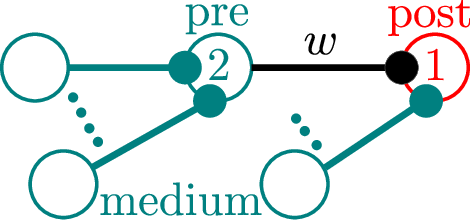}
\par\end{centering}

\caption{We consider the dynamics of a single synapse of synaptic weight $w$
in a mean-field environment.\label{fig:single-synapse-in-MF}}

\end{figure}

Given the mean-field firing frequency $\bar{\lambda}$ for a fixed
mean synaptic weight $\bar{w}$, motivated by the Bethe--Peierls approximation
\cite{huang_statistical_1988}, we consider a single synapse within
such a mean-field environment as illustrated in Fig.~\ref{fig:single-synapse-in-MF}.
 The time-dependence of the synaptic weight is governed by Eq.~\eqref{eq:weight-dynamics}
and the simplification embodied in Eq.~\eqref{eq:simplified stdp}.
With the assumption of Poisson spike trains produced by both the presynaptic
and postsynaptic neurons, the dynamics of the synaptic weight maps
onto an asymmetric random walk (jump process) in $w$-space with $w$-dependent
jump rates and jump sizes \cite{van_kampen_stochastic_2007}. From
Eqs.~\eqref{eq:weight-dynamics} and \eqref{eq:simplified stdp},
the synaptic weight increases by $\Delta\, A_{2}=\Delta\, Y_{2}$
whenever the postsynaptic neuron fires and decreases by $rwB_{1}=rwY_{1}$
whenever the presynaptic neuron fires. As the presynaptic neuron is
driven by the mean-field environment, its firing frequency $\lambda_{2}$
is given by the mean-field firing frequency $\bar{\lambda}$. On the
other hand, the postsynaptic neuron is partly driven by the synapse
in question (Fig.~\ref{fig:single-synapse-in-MF}), thus its firing
frequency $\lambda_{1}$ is $w$ dependent. As calculated from the
neuron response function $\lambda\left(G\right)$, this firing rate
is given by \begin{equation}
\lambda_{1}=\lambda\left(\left[\left(K-1\right)\bar{w}+w\right]\bar{Y}\right),\label{eq:lambda1}\end{equation}
where, even though the synaptic weight $w$ is singled out from among
the $K$ afferent synapses, all of the presynaptic neurons are from
the mean-field environment (see Fig.~\ref{fig:single-synapse-in-MF})
with the active transmitter fraction $\bar{Y}$. As for the step sizes,
since $A_{2}$ depends on the firing of the presynaptic neuron, its
value is given by the mean-field value $\bar{Y}$. On the other hand,
$B_{1}$ is determined by the firing of postsynaptic neuron; therefore
its value is also $w$ dependent and given by $B_{1}=Y\left(\lambda_{1}\right)$.
As mentioned earlier, we set $A=B=Y$ to reduce the number of variables
and computational intensiveness. In general, the system can have different
characteristic functions $A\left(\lambda\right)$ and $B\left(\lambda\right)$
for potentiation and depression.

\subsection{Mean-field synaptic weight}

While the {}``random walk'' of synaptic weight is under the influence
of the mean-field environment, a self-consistency condition requires
the average value of the singled-out synaptic weight from the random
walk process to be the same as the assumed mean-field value $\bar{w}$.
Assuming a normalized stationary synaptic-weight distribution from
the random walk, $P_{s}\left(w\right)$, which is $\bar{w}$-dependent,
the self-consistency requires \begin{equation}
\bar{w}=\left\langle w\right\rangle \equiv\int_{0}^{\infty}wP_{s}\left(w;\bar{w}\right)dw.\label{eq:self-consistent-mean-weight}\end{equation}
The stationary synaptic weight distribution, following the stochastic
dynamics \eqref{eq:simplified stdp}, can not be easily determined
even with the assumption of Poisson spike trains. However, straightforward
numerical simulations of a system of a single synapse and two Poisson
neurons can be used to obtain the stationary distribution $P_{s}\left(w;\bar{w}\right)$
to any reasonable precision. %
\begin{figure}
\begin{centering}
\includegraphics[scale=1.3]{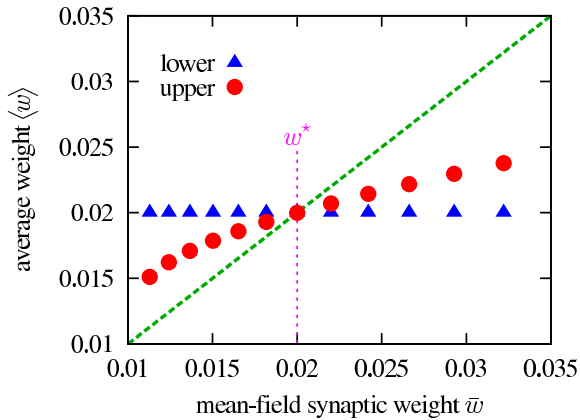}
\par\end{centering}

\caption{Average synaptic weight from the simulation of a single synapse between
two Poisson neurons as shown schematically in Fig.~\ref{fig:single-synapse-in-MF}
versus the mean-field synaptic weight $\bar{w}$ for $w^{\star}=0.02$.
The mean-field firing frequencies $\bar{\lambda}$ used in the single
synapse simulations are given by Fig.~\ref{fig:MF-frequency-on-weight},
and the results for both the upper and lower fixed points (marked
by corresponding symbols in Fig.~\ref{fig:MF-frequency-on-weight})
are plotted.\label{fig:Mean-synaptic-weight}}

\end{figure}
 Figure \ref{fig:Mean-synaptic-weight} shows typical plots of the
average synaptic weight from the random walk process as a function
of the mean-field synaptic weight $\bar{w}$. Notice that the symbols
are lower (higher) than the dashed line on the left (right) meaning
that a deviation of $\bar{w}$ from $w^{\star}$ results in a smaller
deviation of $\left\langle w\right\rangle $ from $w^{\star}$. Hence,
the $\bar{w}=w^{\star}$ is a stable solution for the condition \eqref{eq:self-consistent-mean-weight}.

\subsection{Fokker--Planck equation}

Since the synaptic weight dynamics \eqref{eq:weight-dynamics} {[}or
\eqref{eq:simplified stdp}{]} only depend on the current weight of
the synapse (a Markov process), we can approximate the dynamics of
the weight distribution for small jumps with a master equation ignoring
higher-order moments in the Kramers--Moyal expansion (see, \emph{e.g.},
\cite{van_kampen_stochastic_2007}). A Fokker--Planck equation formalism
has been applied to populations of neurons in previous studies of
the distributions in membrane potential \cite{fourcaud_dynamics_2002,brunel_noise_2006,cateau_relation_2006}.
Here we apply this type of formalism to a population of plastic synapses
similar to the work by Rubin, Lee, and Sompolinsky \cite{rubin_equilibrium_2001}.

Equation \eqref{eq:weight-dynamics} or its simplification \eqref{eq:simplified stdp}
can be viewed as a stochastic equation governing the jumps in synaptic
weight space upon arrival of spikes and which allows us to calculate
the drift \begin{eqnarray}
v & \equiv & \lim_{\Delta t\rightarrow0}\frac{\left\langle w\left(\Delta t\right)-w\right\rangle _{w\left(0\right)=w}}{\Delta t}\nonumber \\
 & = & \Delta\bar{Y}\lambda_{1}-rwY\left(\lambda_{1}\right)\bar{\lambda},\label{eq:drift}\end{eqnarray}
where $\lambda_{1}$ is given by Eq.~\eqref{eq:lambda1}. One also
has for the diffusion coefficient (arising from the second moment
of the synaptic weight changes) \begin{eqnarray}
D & \equiv & \frac{1}{2}\lim_{\Delta t\rightarrow0}\frac{\left\langle \left[w\left(\Delta t\right)-w\right]^{2}\right\rangle _{w\left(0\right)=w}}{\Delta t}\nonumber \\
 & = & \frac{1}{2}\Delta^{2}\bar{Y}^{2}\lambda_{1}+\frac{1}{2}r^{2}w^{2}\left[Y\left(\lambda_{1}\right)\right]^{2}\bar{\lambda}.\label{eq:diffusion}\end{eqnarray}
 The dynamics of the synaptic weight distribution $P\left(w\right)$
is then approximated for small jumps by the Fokker--Planck equation
\begin{equation}
\frac{\partial P\left(w\right)}{\partial t}=-\frac{\partial}{\partial w}\left[vP\left(w\right)\right]+\frac{\partial^{2}}{\partial w^{2}}\left[DP\left(w\right)\right]\label{eq:Fokker-Planck}\end{equation}
when higher-order moments in the Kramers--Moyal expansion are ignored.
Consider the stationary state $P_{s}\left(w\right)$ of the weight
distribution. The Fokker--Planck equation \eqref{eq:Fokker-Planck}
can be integrated once to yield \begin{equation}
vP_{s}\left(w\right)=\frac{\partial}{\partial w}\left[DP_{s}\left(w\right)\right]+\textrm{const.},\end{equation}
which is formally solved by\begin{equation}
P_{s}\left(w\right)\propto\frac{1}{D}e^{\int^{w}dw'v\left(w'\right)/D\left(w'\right)}.\label{eq:formal-solution-stationary-dist}\end{equation}
The peak positions of the distribution \eqref{eq:formal-solution-stationary-dist},
$\hat{w}$, (if they exist) are given by $v\left(\hat{w}\right)=D'\left(\hat{w}\right)$.
Since the diffusion coefficient \eqref{eq:diffusion} is one-order
higher in the plasticity rate $r$ compared to the drift \eqref{eq:drift},
under the small jump approximation, the peak positions are given approximately
by the zeros of $v$. When $\bar{w}=w^{\star}\equiv\Delta/r$, we
can verify that the drift \eqref{eq:drift} is zero and thus the distribution
$P_{s}\left(w\right)$ should peak at $w=\bar{w}$. We can expand
$v$ and $D$ in powers of $w-\bar{w}$ and keep only the first order
terms. The exponent of Eq.~\eqref{eq:formal-solution-stationary-dist}
becomes \begin{equation}
\int^{w}dw'\frac{w^{\star}-w'}{r\bar{Y}{w^{\star}}^{2}}=\frac{-\left(w^{\star}-w\right)^{2}}{2r\bar{Y}{w^{\star}}^{2}}+\mbox{const.},\label{eq:exponent-in-weight-distribution}\end{equation}
which results in a Gaussian distribution for $P_{s}\left(w\right)$.
The distribution is very narrow; the width of the distribution is
given by \begin{equation}
\Delta w\simeq\sqrt{r\bar{Y}}w^{\star}\label{eq:weight-width-const}\end{equation}
which corresponds to only about 1\% of $w^{\star}$ for the parameters
we have considered in Table \ref{tab:Values-of-parameters} above.
Simulations for a fully connected network using the same model and
parameters yield a width about five times larger but with the same
scaling power in $r$ in the noise-dominated regime \cite{chen_simulation-results}.
We will return to this point below.

\subsection{Stability analysis within mean-field theory}

The condition $v\left(\hat{w}\right)=0$ for the peak position $\hat{w}$
of the synaptic-weight distribution gives rise to \begin{equation}
\Delta\bar{Y}\hat{\lambda}_{1}=r\hat{w}Y\left(\hat{\lambda}_{1}\right)\bar{\lambda},\end{equation}
where $\hat{\lambda}_{1}=\lambda_{1}\left(\hat{w}\right)$. Using
the mean-field response function \eqref{eq:synapse-response-function}
for the synapses, we get \begin{equation}
\hat{w}=w^{\star}\frac{1+u\left(\tau_{D}+\tau_{R}\right)\hat{\lambda}_{1}}{1+u\left(\tau_{D}+\tau_{R}\right)\bar{\lambda}}.\label{eq:flow-balance-condition}\end{equation}
 It is straightforward to verify that when the mean-field synaptic
strength is given by $\bar{w}=w^{\star}$, the peak position is given
by $\hat{w}=\bar{w}$, providing a self-consistent solution. To determine
the stability of the solution, we start with a small deviation $\delta\bar{w}$
of $\bar{w}$ from $w^{\star}$ and find the resultant deviation $\delta\hat{w}$
of the peak position $\hat{w}$ from $\bar{w}$. Assuming the neuron
response function $\lambda\left(G\right)$ is smooth around the mean-field
total conductance $\bar{G}=K\bar{w}\bar{Y}$, the firing frequency
of the postsynaptic neuron can be expanded\begin{equation}
\hat{\lambda}_{1}\approx\bar{\lambda}+\lambda'\left(\bar{G}\right)\delta\hat{w}\bar{Y}\end{equation}
where $\lambda'$ is the first derivative of the mean-field response
function $\lambda\left(G\right)$. The mean-field response function
$Y\left(\lambda\right)$ for the active transmitter can also be expanded
as \begin{equation}
Y\left(\lambda_{1}\right)=\bar{Y}\left[1+Y'\left(\bar{\lambda}\right)\lambda'\left(\bar{G}\right)\delta\hat{w}\right].\end{equation}
The condition $v\left(\hat{w}\right)=0$ results in

\begin{equation}
\bar{w}-w^{\star}=\delta\bar{w}=\left\{ w^{\star}\lambda'\left(\bar{G}\right)\left[\frac{\bar{Y}}{\bar{\lambda}}-Y'\left(\bar{\lambda}\right)\right]-1\right\} \delta\hat{w}.\end{equation}
For the $\bar{w}=w^{\star}$ solution to be stable, $\delta\hat{w}$
and $\delta\bar{w}$ should be of opposite signs so that the perturbation
can be damped. Thus, we need \begin{equation}
w^{\star}\lambda'\left(\bar{G}\right)\left[\frac{\bar{Y}}{\bar{\lambda}}-Y'\left(\bar{\lambda}\right)\right]<1,\end{equation}
 or, for given total synaptic conductance $\bar{G}$, \begin{equation}
w^{\star}<w_{\mathrm{b}}^{\star}\left(\bar{G}\right)\equiv\left\{ \lambda'\left(\bar{G}\right)\left[\frac{\bar{Y}}{\bar{\lambda}}-Y'\left(\bar{\lambda}\right)\right]\right\} ^{-1}.\label{eq:stability-condition}\end{equation}
For the integrate-and-fire model we have considered, in the noise-dominated
regime $\lambda=\lambda_{N}$ and $\lambda'=0$, so the condition
\eqref{eq:stability-condition} is satisfied. In the large $\bar{G}$
limit, the firing frequency $\lambda$ increases linearly with $\bar{G}$
while the transmitter fraction $Y$ for the TUM model saturates leading
to a linear $w_{\mathrm{b}}^{\star}\left(\bar{G}\right)$. For self-consistency,
the membrane conductance is given by $\bar{G}=Kw^{\star}\bar{Y}$
for the $\bar{w}=w^{\star}$ solution, and it is thus left to the
constant $K$ (representing the number of connections per neuron)
to determine whether the solution will remain stable. For the range
of the parameters we have considered, the solution $\bar{w}=w^{\star}$
remains stable for large $w^{\star}$ as long as $K>1$. However,
it is straightforward to verify from Eq.~\eqref{eq:stability-condition}
that $\bar{w}=w^{\star}$ can not remain a stable solution at large
$w^{\star}$ for models in which, for example, $\lambda\left(G\right)$
saturates for large $G$ or $Y\left(\lambda\right)$ increases linearly
(when it no longer represents a fraction) with $\lambda$ for large
$\lambda$. Under these situations, this analysis suggests that one
may find runaways or pile-ups of the synaptic weights in the system.

\section{Correction to the mean-field theory\label{sec:Correction-to-MFT}}

In the mean-field considerations, we characterized all neurons with
a single firing frequency $\bar{\lambda}$, all synapses with a constant
active transmitter fraction $\bar{Y}$, and the {}``network'' by
a constant strength $\bar{w}$. This is an over simplification even
within the mean-field approach. We retain the characterization of
the environment by a single $\bar{w}$. However, a neuron within a
network experiences, instead of a constant total synaptic conductance,
the bombardment of synaptic conductance pulses issuing from the spike
trains of corresponding presynaptic neurons. This shot-noise-like
fluctuation is most prominent when the amplitude of the pulses is
comparable with the mean of the total synaptic conductance. When the
number $K$ of afferent synapses per neuron and the total frequency
$\lambda_{\mathrm{total}}$ of presynaptic events are small, the mean
level of the total synaptic conductance is comparable with the amplitude
of each individual pulse. Consequently, the assumption of a constant
total synaptic conductance will be inadequate.

In the mean-field approximation above, we also assumed that the neurons
fire following Poisson statistics. However, when integrate-and-fire
neurons are driven by a constant total synaptic conductance (as in
the limit of large $K$ and $\lambda_{\mathrm{total}}$), the time
between threshold crossings will also be constant, and the resulting
spike trains of the neuron will be periodic. Combined with the Poisson
external noise, the intervals between firings will follow a Poisson
distribution only up to the period of threshold crossings, where there
will be a $\delta$-function representing the periodic firings. Nonetheless,
Poisson and periodic firings result in the same mean level of active
transmitter fraction in the limits of low and high firing frequencies,
while the corrections in the intermediate frequency regime remains
insignificant for the TUM model that we considered (see the dashed
line in Fig.~\ref{fig:Firing-frequency-on-G}). We will thus retain
the assumption of Poisson firing for the following discussion.

\subsection{Effect of fluctuation in total synaptic conductance}

Apart from making an analytically tractable system, a mean-field approach
does not necessarily require approximating the total synaptic conductance
as a constant in time. Here, we consider the total synaptic conductance
$G$ as a time dependent function and model the dynamics of $G$ approximately
with \begin{equation}
\frac{dG}{dt}=-\frac{G}{\tau_{D}}+JS_{\mathrm{total}},\label{eq:total_conductance_dynamics}\end{equation}
where $S_{\mathrm{total}}$ is a Poisson spike train with the frequency
$\lambda_{\mathrm{total}}$ accounting for any presynaptic events,
and the jump size $J$ is a constant modeling the increase of the
total synaptic conductance due to each spike in $S_{\mathrm{total}}$.
(In the spirit of mean-field theory, we simplify the dynamics of $G$
by assuming a constant jump size $J$. In general, the jump size of
the total synaptic conductance for each presynaptic event is a random
variable depending on the active transmitter fraction of the firing
presynaptic neuron.) The average total synaptic conductance $\bar{G}$
is related to frequency $\lambda_{\mathrm{total}}$ and jump size
$J$ by \begin{equation}
\bar{G}=\lambda_{\mathrm{total}}\tau_{D}J,\label{eq:total_cond-on-freq-J}\end{equation}
and the ensemble of $G\left(t\right)$ can be characterized by two
independent parameters from the set $\lambda_{\mathrm{total}}$, $J$,
and $\bar{G}$. Here, we choose $\bar{G}$ and $J$ to characterize
the ensemble, and the resulting ensemble-averaged firing frequency
$\lambda_{\mathrm{out}}\left(\bar{G},J\right)$ of the post-synaptic
neuron is a generalization of the response function $\lambda\left(\bar{G}\right)$
considered in Section~\ref{sec:Mean-field-approximation}.

To evaluate $\lambda_{\mathrm{out}}\left(\bar{G},J\right)$ for given
values of $\bar{G}$ and $J$, we simulate the dynamics \eqref{eq:total_conductance_dynamics}
with a Poisson spike train of frequency $\lambda_{\mathrm{total}}$
inferred from Eq.~\eqref{eq:total_cond-on-freq-J} to obtain the
time-dependent total synaptic conductance $G\left(t\right)$, which
we use in the dynamics of the membrane potential \eqref{eq:membrane-with-conductance};
the measured average firing frequency of the postsynaptic neuron gives
us the value of $\lambda_{\mathrm{out}}\left(\bar{G},J\right)$. %
\begin{figure}
\begin{centering}
\includegraphics[scale=1.1]{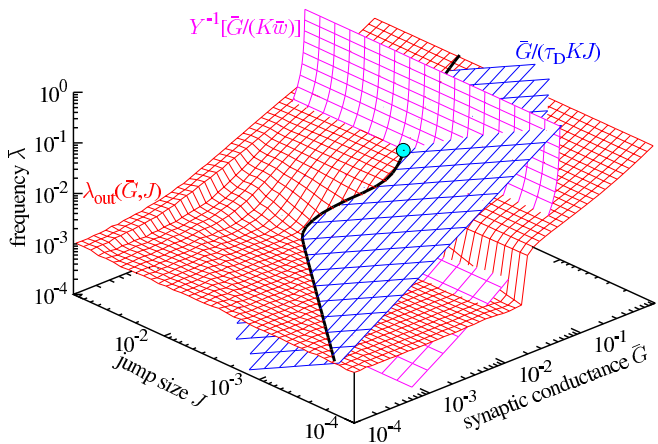}
\par\end{centering}

\caption{Extended space of Fig.~\ref{fig:Firing-frequency-on-G} when the
fluctuation in synaptic conductance $G$ is considered (represented
by the jump size $J$). The extended neuron response function \eqref{eq:extended-neuron-response},
the synapse response function $Y\left(\bar{\lambda}\right)=\bar{G}/\left(K\bar{w}\right)$,
and the relation \eqref{eq:network-relation-4-cond-J-lambda} determine
the stationary states of a mean-field network (marked by a circle
on the surfaces).\label{fig:intersection-in-3d}}

\end{figure}
 The process is repeated to generate the surface \begin{equation}
\bar{\lambda}=\lambda_{\mathrm{out}}\left(\bar{G},J\right)\label{eq:extended-neuron-response}\end{equation}
as plotted in Fig.~\ref{fig:intersection-in-3d}; compare to the
line representing $\lambda\left(G\right)$ as plotted in Fig.~\ref{fig:Firing-frequency-on-G}.
The jump size $J$ is an estimator of the fluctuations in the total
synaptic conductance $G\left(t\right)$, and in the $J=0$ limit,
the two-parameter response function $\lambda_{\mathrm{out}}\left(\bar{G},\lambda\right)$
reduces to the single parameter response function $\lambda_{\mathrm{out}}\left(\bar{G},0\right)=\lambda\left(\bar{G}\right)$.

As we are considering a mean-field network with $K$ afferent synapses
per neuron, the total frequency of presynaptic events for a postsynaptic
neuron is given by $\lambda_{\mathrm{total}}=K\bar{\lambda}$, where
$\bar{\lambda}$ is the mean-field firing frequency for any neuron.
Combining with the relation \eqref{eq:total_cond-on-freq-J}, this
gives us the condition \begin{equation}
\frac{\bar{G}}{K\tau_{D}J}=\bar{\lambda},\label{eq:network-relation-4-cond-J-lambda}\end{equation}
as plotted in Fig.~\ref{fig:intersection-in-3d}, forming a planar
surface in the logarithmic-scale $\bar{G}$--$J$--$\bar{\lambda}$
space.

Whatever the value of $J$, the synaptic response function $Y\left(\bar{\lambda}\right)$
determines the mean total synaptic conductance $\bar{G}=K\bar{w}Y\left(\bar{\lambda}\right)$
when $\bar{\lambda}$ is given for the mean-field network. The inverse
\begin{equation}
\bar{\lambda}=Y^{-1}\left(\frac{\bar{G}}{K\bar{w}}\right),\label{eq:inverse-synaptic-response}\end{equation}
also plotted in Fig.~\ref{fig:intersection-in-3d} with Eq.~\eqref{eq:transmitter-to-conductance},
represents the convex surface invariant along the $J$-axis. Along
with the surfaces \eqref{eq:extended-neuron-response} and \eqref{eq:network-relation-4-cond-J-lambda},
the intersections of the three surfaces determine the fixed points
of the system (marked with a circle in Fig.~\ref{fig:intersection-in-3d}).
While the neuron response function $\lambda_{\mathrm{out}}\left(\bar{G},J\right)$
is evaluated numerically, we determine the fixed points in the mean-field
firing frequency numerically from the intersections of the three surfaces
in Fig.~\ref{fig:intersection-in-3d}, varying the mean synaptic
weight $\bar{w}$ for a given number of afferent synapses $K$ per
neuron. %
\begin{figure}
\begin{centering}
\includegraphics[scale=1.3]{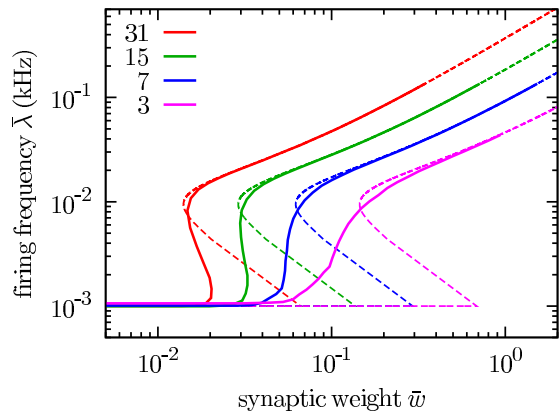}
\par\end{centering}

\caption{Numerical mean-field phase diagram (firing frequency $\bar{\lambda}$
versus synaptic weight $\bar{w}$) for various numbers $K$ (as labeled
in legend) of afferent synapses per neuron under fluctuating total
synaptic conductance (drawn lines) compared with the analytical results
when the total synaptic conductance $G=\bar{G}$ is assumed to be
constant (dashed line).\label{fig:Numerical-mean-field-phase}}

\end{figure}
 The resulting phase diagram is plotted in Fig.~\ref{fig:Numerical-mean-field-phase}
for various $K$ values. The hysteresis vanishes at $K=K_{C}\approx10$,
and the transition from noise dominance to persistent activity becomes
continuous. In our case, the fluctuation comes from the fact that
the number of afferent synapses is finite. We expect that, in a real
biological network, other sources of fluctuations can play a similar
role to smooth out the first-order phase transition separating a noise
dominated regime from persistent activity as predicted by the analytical
mean-field theory of Section \ref{sec:Mean-field-approximation}.

\subsection{Correction to synaptic weight distribution}

Besides smoothing out the first-order phase transition in a mean-field
network, the fluctuation in total synaptic conductance can also influence
the stationary distribution of synaptic weights. We have noted that
the width of the synaptic weight distribution predicted by the simplest
mean-field approach (about 1\% of $w^{\star}$) is too small in comparison
with results of simulations on fully connected plastic integrate-and-fire
networks (about 5\% of $w^{\star}$ \cite{chen_simulation-results}).
One of the approximations we made in arriving at Eq.~\eqref{eq:exponent-in-weight-distribution}
is to replace the time-dependent $Y$ (or the fractions $A$ and $B$
that it represents) with its average value. Such an approximation
can be improved by considering the second moment of $Y$ in deriving
the diffusion coefficient $D$ in Eq.~\eqref{eq:diffusion}. We expect
the fluctuation in $Y$ to be important when the mean-field firing
frequency is low. In this limit the time-dependent $Y\left(t\right)$
is a sum of pulses \begin{equation}
Y\left(t\right)=u\sum_{i}\theta\left(t-t_{i}\right)e^{-\left(t-t_{i}\right)/\tau_{D}},\end{equation}
where $\theta\left(t\right)$ is a step function with $\theta\left(t\geq0\right)=1$
and $\theta\left(t<0\right)=0$, $\{t_{i}\}$ are the firing times
of the neuron, and mean $\left\langle Y\right\rangle =u\tau_{D}\lambda$.
When the overlap of the pulses can be ignored, for example, in the
noise dominated regime, the mean square of $Y$ is given by $\left\langle Y^{2}\right\rangle =u^{2}\tau_{D}\lambda/2$,
which should replace the squares of $Y$ in the diffusion coefficient
\eqref{eq:diffusion}. This leads to a width of the synaptic weight
distribution of \begin{equation}
\Delta w\simeq\sqrt{r\frac{\left\langle Y^{2}\right\rangle }{\bar{Y}}}w^{\star}=\sqrt{ur/2}w^{\star}\label{eq:weight-width-fluc}\end{equation}
instead of Eq.~\eqref{eq:weight-width-const}. For the values of
the parameters we have considered, this represents about 5\% of $w^{\star}$
and is similar to the simulation results from a fully connected network
\cite{chen_simulation-results} in the noise-dominated regime. We
have verified the analytical result \eqref{eq:weight-width-fluc}
through numerical simulations of a random walk process for the synaptic
weight described by the dynamics \eqref{eq:simplified stdp}. %
\begin{figure}
\begin{centering}
\includegraphics[scale=1.3]{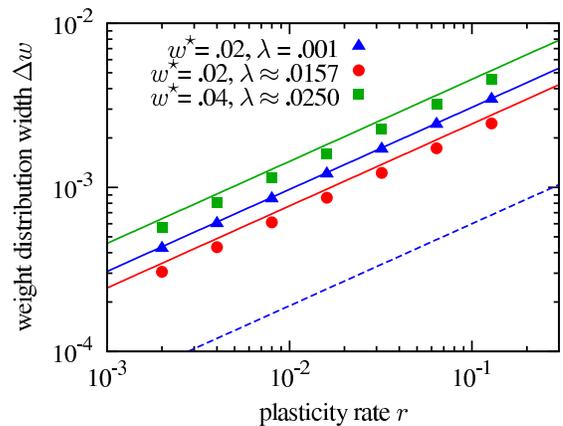}
\par\end{centering}

\caption{Numerical results (symbols) for the width of synaptic-weight distribution,
$\Delta w$, as a function of plasticity rate, $r$, in a limited
system of a single synapse between two neurons with Poisson spike
trains following the dynamics \eqref{eq:simplified stdp} and operating
at three of the fixed points predicted by the mean-field theory (as
marked by corresponding symbols in Fig.~\ref{fig:MF-frequency-on-weight}).
The results are compared with the analytical predictions (lines of
corresponding colors) from Eq.~\eqref{eq:weight-width-fluc} (drawn
lines) and Eq.~\eqref{eq:weight-width-const} (dashed line).\label{fig:limited-system-results}}

\end{figure}
 As shown in Fig.~\ref{fig:limited-system-results}, the scaling
in plasticity rate $r$ continues to be described by $\Delta w\sim r^{\alpha}$
with $\alpha\simeq1/2$ in all cases, while the amplitudes match well
for the noise-dominated fixed point but show small deviations when
the system is persistently active. In general, we can expect fluctuations
to play an important role in determining the synaptic weight distribution
of a plastic network. Here we have shown that a significant correction
can be accounted for by considering the fluctuation in the neural
transmission for a limited system of a single synapse between two
neurons with Poisson spike trains. Even without an analytical solution,
such a system can be easily simulated to any desirable accuracy to
obtain the stationary synaptic weight distribution as well as the
average synaptic weight $\left\langle w\right\rangle $, which provides
the self-consistency condition \eqref{eq:self-consistent-mean-weight}
for the mean-field theory.

Here we have preserved the basic structure of the mean-field approach
under the additional consideration of fluctuations in synaptic conductance
by extending the neuron and synapse response functions to functions
of two variables. The additional degree of freedom, the average jump
size $J$ of active transmitter fractions, was fixed by the condition
\eqref{eq:network-relation-4-cond-J-lambda} coming from the nature
of such a fluctuation and does not enter the final result \eqref{eq:weight-width-fluc}
explicitly.

\section{Summary and conclusion\label{sec:Summary-and-conclusion}}

Models of plastic neural networks can generally be broken down into
three parts: (i) the modeling of the dynamics of the neuron state
(represented by the membrane potential), (ii) the modeling of synaptic
transmission, and (iii) the plasticity rules. We have applied a mean-field
theory that follows this basic structure. In the mean-field framework
of the current study, we reduce the dynamics of the neuron state to
a response function $\lambda\left(G\right)$ representing the mean
firing frequency of a neuron when it is driven by a constant total
synaptic conductance $G$. Similarly, the dynamics of synaptic transmission
is also reduced to a characteristic response function $Y\left(\lambda\right)$
representing the mean fraction of active neural transmitter given
the Poisson firing frequency $\lambda$ of the presynaptic neuron.
With the mean-field synaptic weight $\bar{w}$ of a network with $K$
afferent synapses per neuron and the mean-field firing frequency $\bar{\lambda}$,
the average total synaptic conductance is given by $\bar{G}=K\bar{w}Y\left(\bar{\lambda}\right)$.
This allows one to plot the synapse response function $Y\left(\lambda\right)$
along with the neuron response function $\lambda\left(G\right)$ (see
Fig.~\ref{fig:Firing-frequency-on-G}) to determine self-consistently
the mean-field firing frequency $\bar{\lambda}$ as a function of
$\bar{w}$ and $K$.

The mean-field firing frequency $\bar{\lambda}$ and the mean synaptic
weight $\bar{w}$ characterize the mean-field network completely.
We then use this mean-field network as an environment and investigate
the dynamics of spike-timing-dependent plasticity \eqref{eq:weight-dynamics}
of a single synapse in such an environment. Assuming Poisson statistics
for the spike trains, the dynamics \eqref{eq:weight-dynamics} can
be viewed as describing a type of random walk in synaptic weight space,
where the frequency of potentiation and the jump size of depression
are dependent on the weight of this synapse. The distribution of synaptic
weights can be calculated numerically for the stationary state using
straightforward simulations of the single synapse system. Under the
approximation of small jumps, the stationary distribution can be calculated
analytically from the Fokker--Planck equation arising from the master
equation. The self-consistency of the mean-field approach is completed
by requiring that the mean of the synaptic weight distribution $\left\langle w\right\rangle $
reproduces the mean-field synaptic weight $\bar{w}$ of the environment.

We have considered a network of integrate-and-fire neurons \cite{dayan_theoretical_2001}
coupled through synapses with TUM dynamics \cite{tsodyks_synchrony_2000}
within the mean-field framework outlined above. We chose integrate-and-fire
neurons for their simplicity and wide use. We chose the TUM model
of neural transmission for its features and flexibility as noted above.
On the simplest level for a static network, our mean-field approach
amounts to finding intersects of the neuron response function and
synapse response function, each of which can be calculated separately
from the particular neuron model and synapse model selected. Our choice
of model happens to allow us to find analytical forms of the mean-field
response functions for the neurons and synapses, and predict a first-order
phase transition from a noise-dominated regime to a regime of persistent
activity as the mean-field synaptic weight is increased. (However,
see below.) In general, these response functions can be computed numerically
in a straightforward fashion for a variety of neuron and synapse models
regardless of the number of empirical parameters they might carry.
In Fig.~\ref{fig:Firing-frequency-on-G}, we showed the results of
such calculation for the Morris--Lecar neuron used in \cite{volman_calcium_2007}
which was defined by more than twenty parameters.

For the dynamics of the synaptic weight, for specificity, we follow
the spike-timing-dependent plasticity rules proposed by van Rossum,
Bi, and Turrigiano \cite{van_rossum_stable_2000}, with additive potentiation
and multiplicative depression. The Fokker--Planck analysis of the
corresponding random walk process predicts a narrow Gaussian distribution
for the synaptic weight centered around $w^{\star}$, the control
parameter \eqref{eq:wstar-definition} entering the plasticity rules.
{[}See, \emph{e.g.}, Eq.~\eqref{eq:simplified stdp}.{]} We apply
a small perturbation to the $\left\langle w\right\rangle =\bar{w}=w^{\star}$
solution, analyze its stability analytically and numerically, and
find it to be stable for any number of afferent synapse per neuron
$K\geq1$ for the model and parameter ranges we considered. However,
the analytical expression \eqref{eq:stability-condition} for the
stability, which is generic for any neuron and synapse response functions,
also suggests that for models in which the neuron firing rate $\lambda\left(G\right)$
saturates for large conductance $G$ (for example, models with a refractory
period for the firing of neurons) or where the synapse response function
$Y\left(\lambda\right)$ does not saturate for large $\lambda$, (for
example, when the effect of presynaptic firings is additive and $Y$
no longer represents a fraction), the fixed point $\bar{w}=w^{\star}$
can not remain stable for large $w^{\star}$. It is then possible
to have runaway or pile-up in the resulting synaptic weight distribution.

For a network of finite $K$ and low overall frequency $\lambda_{\mathrm{total}}$
of presynaptic events, the {}``shot noise'' due to the discrete
nature of presynaptic spikes is not negligible, and one at least needs
to expand the description of the mean-field response of a neuron to
include temporal fluctuations. We have done this approximately via
a two-variable function, \emph{e.g.}, $\lambda_{\mathrm{out}}\left(\bar{G},J\right)$,
where $J$, representing fluctuations, describes the size of the jumps
in a neuron's total synaptic conductance for each presynaptic event.
We evaluate this two-variable response function numerically for a
single neuron, modeling its total synaptic conductance with a simple
stochastic jump-and-decay process \eqref{eq:total_conductance_dynamics}.
The results suggest the disappearance of the first-order transition
when the number, $K$, of afferent synapses per neuron is less than
a critical value $K_{C}\approx10$. In realistic situations fluctuations
can be expected to smear out the first-order transition. When a corresponding
fluctuation is considered for the variable $Y$ governing jump sizes
in the plasticity rules \eqref{eq:simplified stdp}, the Fokker--Planck
approach predicts a broader Gaussian distribution which has a width
similar to the observed width in full network simulations \cite{chen_simulation-results}.
The results of extensive simulations on a fully connected plastic
network will be published elsewhere.

The inclusion of fluctuations at some level, such as within our extended
mean-field theory (see Sec. \ref{sec:Correction-to-MFT}), hints at
a possible critical state of the system at the endpoint of a first-order
transition line in analogy with the vapor pressure curve of a fluid,
see, e.g., \cite{huang_statistical_1988}. While criticality in neural
\emph{avalanches} has been observed by Beggs and Plenz \cite{beggs_neuronal_2003},
within the extended mean-field analysis employed in the current study
the system does not appear to organize into such a critical state
without a requisite tuning of the fluctuation-amplitude and perhaps
plasticity parameters, \emph{e.g.}, $w^{\star}$. This is in contrast
with the suggestion that such criticality should be self-organized
\cite{beggs_neuronal_2003,chialvo_critical_2004}. It thus will be
of great interest to find missing elements, possibly in the plasticity
rules, that could dynamically push the network close to a critical
state. Along this direction, one can expect a variety of model candidates,
which can be easily subjected to the type of mean-field scheme outlined
in the current study to provide additional qualitative and semi-quantitative
insight into their plausibility. We also note in passing that {}``all-to-all''
network simulations using the plasticity rules and neuron modeling
described in this paper have not revealed evidence for a self-organized
critical state. In that regard, sparse networks would appear to be
better suited, but mean-field analysis such as presented here will
mainly have only qualitative use. Our simulations of integrate-and-fire
and other neuronal networks with the plasticity rules used here will
be presented elsewhere.

As noted in the Introduction, the current study cannot begin to address
how brains function or form. As a real brain is never uniform or stationary,
we do not expect the model systems presented here to address or reproduce
its dynamics. However, there are cultured networks consisting of hundreds
of neurons with virtually {}``all-to-all'' interactions \cite{bi_synaptic_1998,segev_observations_2001,shefi_morphological_2002,beggs_neuronal_2003,lai_growth_2006}.
The studies of these networks \emph{in vitro} is an important steppingstone
towards the understanding of more complicated networks. The dynamics
of these cultured network are on a scale very much accessible to current
neural network modeling, such as the one presented here, and simulation,
as noted, to be presented elsewhere.
\begin{acknowledgments}
We are grateful for helpful discussions and insights from Profs. Guo-Qiang
Bi, Jonathan E. Rubin, G. Bard Ermentrout, and Dr. Richard C. Gerkin
during the course of our work. This research was supported in part
by Computational Resources on PittGrid (www.pittgrid.pitt.edu).
\end{acknowledgments}
\bibliographystyle{apsrev}
\bibliography{mf-ifn-add,mf-ifn}

\end{document}